# Status of the GAMMA-400 Project

A.M. Galper([1])([2]), O. Adriani([3]), R.L. Aptekar([4]), I.V. Arkhangelskaja([2]),
A.I. Arkhangelskiy([2]), M. Boezio([5]), V. Bonvicini([5]), K.A. Boyarchuk([6]), Yu.V. Gusakov([1]),
M.O. Farber([2]), M.I. Fradkin([1]), V.A. Kachanov([7]), V.A. Kaplin([2]), M.D. Kheymits([2]),
A.A. Leonov([2]), F. Longo([5]), P. Maestro([8]), P. Marrocchesi([8]), E.P. Mazets([4]), E. Mocchiutti([5]),
A.A. Moiseev([9]), N. Mori([3]), I. Moskalenko([10]), P.Yu. Naumov([2]), P. Papini([3]), P. Picozza([11]),
V.G. Rodin([12]), M.F. Runtso([2]), R. Sparvoli([11]), P. Spillantini([11]), S.I. Suchkov([1]),
M. Tavani([13]), N.P. Topchiev([1]), A. Vacchi([5]), E. Vannuccini([3]), Yu.T. Yurkin([2]), N. Zampa([5]),
and V.G. Zverev([2])

([1]) Lebedev Physical Institute, Russian Academy of Sciences - Leninskii pr. 53, RU-119991 Moscow, Russia;
([2]) National Research Nuclear University MEPhI - Kashirskoe sh. 31, RU-115409 Moscow, Russia;
([3]) Istituto Nazionale di Fisica Nucleare, Sezione di Firenze and Physics Department of University of Florence - Via Sansone 1, I-50019 Sesto Fiorentino (Firenze), Italy;
([4]) Ioffe Physical Technical Institute, Russian Academy of Sciences – ul. Polytekhnicheskaya 26, RU-194021 St. Petersburg, Russia;
([5]) Istituto Nazionale di Fisica Nucleare, Sezione di Trieste - Padriciano 99, I-34012 Trieste, Italy
([6]) Open Joint Stock Company "Research Institute for Electromechanics" -ul. Panfilova 11, RU-143502 Istra, Moscow region, Russia;
([7]) Institute for High Energy Physics – ul. Pobedy 1, RU-142281 Protvino, Moscow region, Russia;
([8]) Istituto Nazionale di Fisica Nucleare, Sezione di Pisa and Physics Department of University os Siena, Via Roma 56, I-53100 Siena, Italy;
([9]) NASA Goddard Space Flight Center and CRESST/University of Maryland, Greenbelt, Maryland 20771, USA;
([10]) Hansen Experimental Physics Laboratory and Kavli Institute for Particle Astrophysics and Cosmology, Stanford University, Stanford, CA 94305, USA;
([11]) Istituto Nazionale di Fisica Nucleare, Sezione di Roma 2 and Physics Department of University of Rome "Tor Vergata" -Via della Ricerca Scientifica 1, I-00133 Rome,Italy;
([12])Space Research Institute, Russian Academy of Sciences – ul. Profsoyuznaya 84/32, RU-117997 Moscow, Russia;
([13]) Istituto Nazionale di Astrofisica – IASF and Physics Department of University of Rome "Tor Vergata", Via della Ricerca Scientifica 1, I-00133 Rome, Italy.

**Abstract**
The preliminary design of the new space gamma-ray telescope GAMMA-400 for the energy range 100 MeV – 3 TeV is presented. The angular resolution of the instrument, 1-2° at $E_\gamma$ ~100 MeV and ~0.01° at $E_\gamma > 100$ GeV, its energy resolution ~1% at $E_\gamma > 100$ GeV, and the proton rejection factor ~$10^6$ are optimized to address a broad range of science topics, such as search for signatures of dark matter, studies of Galactic and extragalactic gamma-ray sources, Galactic and extragalactic diffuse emission, gamma-ray bursts, as well as high-precision measurements of spectra of cosmic-ray electrons, positrons, and nuclei.

## 1. Introduction

One hundred years after Hess' discovery of the radiation coming from space, the main questions about the origin of cosmic rays (CRs), their acceleration, and propagation in the Galaxy and intergalactic space remain open. Energetic CR particles when interacting with matter and radiation fields produce gamma-ray emission and, therefore, studies of gamma-ray sources and diffuse emission provide important clues to the origin and propagation of cosmic rays. A large number of outstanding problems in physics and astrophysics are connected with studies of CRs and associated gamma-ray emission. Among the most pressing issues are the nature of the dark matter, particle acceleration in Galactic and cosmological shocks, the origin of extragalactic diffuse emission, CRs in other galaxies and the role they play in galactic evolution, studies of the Galactic Center and local Galactic environment, CR propagation in the heliosphere, as well as many others.

The last years were rich in new breakthroughs and discoveries thanks to the combined efforts of many very successful space missions, such as PAMELA (Picozza, Galper et al., 2007), AGILE (Tavani et al., 2009), and Fermi-LAT (Atwood et al., 2009); balloon-borne missions, such as ATIC (Guzik et al., 2004), CREAM (Seo et al., 2004), and BESS (Yamamoto et al., 2007); as well as ground-based gamma-ray telescopes, such as MILAGRO (Atkins et al., 2004), H.E.S.S. (Aharonian et al., 2004), VERITAS (Weekes et al., 2010), and MAGIC (Aleksić et al., 2010). Among them are the new measurements of CR electrons-positrons by PAMELA (Adriani et al., 2009), Fermi-LAT (Abdo et al., 2009a), H.E.S.S. (Aharonian et al., 2009), and ATIC (Chang et al, 2008); antiprotons by BESS (Mitchell et al., 2005) and PAMELA (Adriani et al., 2011a); and a discovery of the break in the spectra of CR nuclei by ATIC (Panov et al., 2009), CREAM (Yoon et al., 2011), and PAMELA (Adriani et al., 2011b). In gamma rays - spectra of diffuse Galactic and extragalactic emission by EGRET (Hunter et al., 1997, Sreekumar et al., 1998) and Fermi-LAT (Abdo et al., 2009b); 270 Galactic and extragalactic gamma-ray sources with energies up to 20 GeV by EGRET (Hartman et al., 1999), approximately 50 sources up to 50 GeV by AGILE (Pittori et al., 2009), and nearly 1850 sources up to 100 GeV by Fermi-LAT (Abdo et al., 2011); about a hundred sources of TeV photons by H.E.S.S. (Tibolla et al., 2009), VERITAS (Holder et al., 2011), MAGIC (Cortina et al., 2011), and MILAGRO (Smith et al., 2010).These exciting recent developments show that we have just touched the tip of the iceberg and even more exciting discoveries are still awaiting us.

Further observations with improved angular and energy resolutions in high-energy gamma-ray range are necessary to address the following issues: study unidentified sources which are one third of discovered gamma-ray sources (Abdo et al., 2011), including detailed investigation of the Galactic center, study sources for the energies more than several tens GeV, where the energy ranges of Fermi and ground-based gamma-ray telescopes do not overlap (Abdo et al., 2009c), study variability of many sources, study diffuse gamma-ray emission for energies more than several tens GeV, search for high-energy emission from gamma-ray bursts and transient gamma-ray sources. The largely unexplored energy range of several tens MeV – few hundred MeV with only handful of observed sources has an enormous discovery potential.

The diffuse gamma-ray emission above 100 GeV remains unexplored and may provide some clues to the origin of the dark matter (Abazajian et al. 2011, Beischer et al., 2009). A search for a fine structure in the spectra of high-energy CR electrons-positrons and nuclei is another important direction to pursue.

## 2. The GAMMA-400 gamma-ray telescope
### 2.1. The GAMMA-400 performance

GAMMA-400 is designed to address scientific challenges and will detect gamma rays in the energy range 100 MeV – 3 TeV with an angular resolution of 1-2° at $E_\gamma \sim$100 MeV and ~0.01° at $E_\gamma >$ 100 GeV, an energy resolution of ~1% at $E_\gamma >$ 100 GeV, and an effective area of ~4000 cm$^2$ at $E_\gamma =$ 100 GeV. The instrument will also be used to detect electrons (positrons) in the energy range 1-3000 GeV with a proton rejection factor ~$10^6$ and nuclei in the energy range 250 GeV/nucleon to $10^{15}$ eV/nucleon. The total mass of the instrument is 2600 kg with power consumption of ~2000 W and a telemetry downlink capability of 100 Gbyte/day. Together with GAMMA-400, the space observatory will include the KONUS-FG gamma-ray burst monitor (Aptekar et al., 2009) and a star sensor for determining the GAMMA-400 axes with accuracy of approximately 5".

The GAMMA-400 physical scheme is shown in Fig. 1. The basic idea of the instrument was outlined in earlier publications (Dogiel et al., 1988, Ginzburg et al., 2007, 2009a, 2009b), but the design is being improved (Galper et al., 2011a, 2011b, 2011c) using the results of the current space missions Fermi-LAT and AGILE. Gamma-400 consists of scintillation anticoincidence top and lateral detectors (AC), converter-tracker (C) with approximately 25 layers of double (x, y) silicon strip coordinate detectors (pitch 0.1 mm) interleaved with tungsten conversion foils, scintillation detectors (S1 and S2) of time-of-flight system (TOF), silicon strip coordinate detector CD1 (pitch 0.1 mm), scintillation detectors S3 and S4, silicon arrays (silicon pad detectors with 1×1 $cm^2$ pixels), lateral detectors (LD) from the same Si arrays and tungsten planes, and calorimeter from two parts (CC1 and CC2). The imaging calorimeter CC1 consists of 4 layers of double (x, y) silicon strip coordinate detectors (pitch 0.5 mm) interleaved with tungsten planes, and the electromagnetic calorimeter CC2 consists of BGO crystals. The total converter-tracker thickness is ~1 $X_0$ ($X_0$ is the radiation length). The thickness of CC1 and CC2 is 3 $X_0$ and 22 $X_0$, respectively. The total calorimeter thickness is 25 $X_0$ and 1.2 $\lambda_0$ ($\lambda_0$ is nuclear interaction length) in the vertical direction and 70 $X_0$ and 3.5 $\lambda_0$ in the lateral direction.

Table 1 shows a comparison of basic parameters of space-based and ground-based instruments: EGRET (Thompson et al., 1993), AGILE (Tavani et al., 2009), Fermi (Atwood et al., 2009), CALET (Torii et al., 2008), H.E.S.S. (Aharonian et al., 2007), MAGIC (Aleksić et al., 2010), VERITAS (Weekes et al., 2010), and CTA (The CTA Consortium, 2010). As seen from the table, GAMMA-400 will have superior angular and energy resolutions.

## 2.2. Particle detection

The gamma rays are detected through the conversion into the electron-positron pairs in the converter-tracker. The time-of-flight system, where detectors S1 and S2 are separated by approximately 500 mm, determines the direction of the arriving particle. Silicon strip coordinate detector together with the anticoincidence detectors located at front and sides of the converter-tracker helps to identify gamma rays. The electromagnetic cascade initiated by the electron-positron pair develops in CC1 and CC2 parts of the calorimeter. Additional scintillation detectors S3 and S4 can detect cascade particles.

When measuring from the top-down direction, we use the two main triggering systems: (i) for gamma rays if there is no AC signal, (ii) for electrons (positrons) and nuclei when the AC signal is present.

Electrons (positrons) and nuclei are also detected in the calorimeter when coming from the four lateral directions. In the latter case, we use the four lateral detectors. Silicon arrays (at the top and at the middle) are used to determine the particle charge.

Using thick calorimeter (~25 $X_0$) allows us to extend the energy range up to several TeV and to reach the energy resolution up to ~1% above 100 GeV. The improvement in the angular resolution is achieved by using CC1, which allows us to reconstruct the axis of the cascade, together with the coordinates of the conversion point in multilayer converter. This method allows us to reach the superior angular resolution of ~0.01° above 100 GeV.

The calorimeter scheme together with additional data from other detectors provides proton rejection factor up to ~$10^6$. To increase the instrument efficiency at high energies and to reduce the dead time of the telescope due to the backsplash particles we use the temporal and segmentation methods.

## 3. Spacecraft

The GAMMA-400 space observatory will be installed on the Navigator space service platform produced by Lavochkin Research and Production Association. Two variants of orbit are possible: Lagrange point $L_2$ and high-elliptical orbit. The initial high-elliptical orbital parameters are: an apogee of 300 000 km, a perigee of 500 km, and an inclination of 51.8°. The orbit period will be 7 days. After approximately 230 days GAMMA-400 will leave the Earth's radiation belts and the orbit will change from highly elliptical to approximately circular with median altitude of ~150 000 km.

We are planning to use three basic modes of observations:

- All-sky gamma-ray monitoring in order to search for new sources and to monitor the discovered variable sources;
- Long-term monitoring of selected point sources;
- Observations of the GeV emission from gamma-ray bursts and solar flares using a trigger from the KONUS-FG gamma-ray burst monitor. A specific pointing can also be triggered by other spacecraft as well as by ground-based telescopes.

## 4. Conclusion

At present, we consider a possibility to extend the GAMMA-400 energy range down to approximately 30 MeV. The launch of the GAMMA-400 space observatory is planned in 2018. The expected mission duration is longer than 7 years.

## Acknowledgements

This work was supported by Space Council of the Russian Academy of Sciences and the Russian Space Agency.

**Figure caption**
Fig. 1. The GAMMA-400 physical scheme.

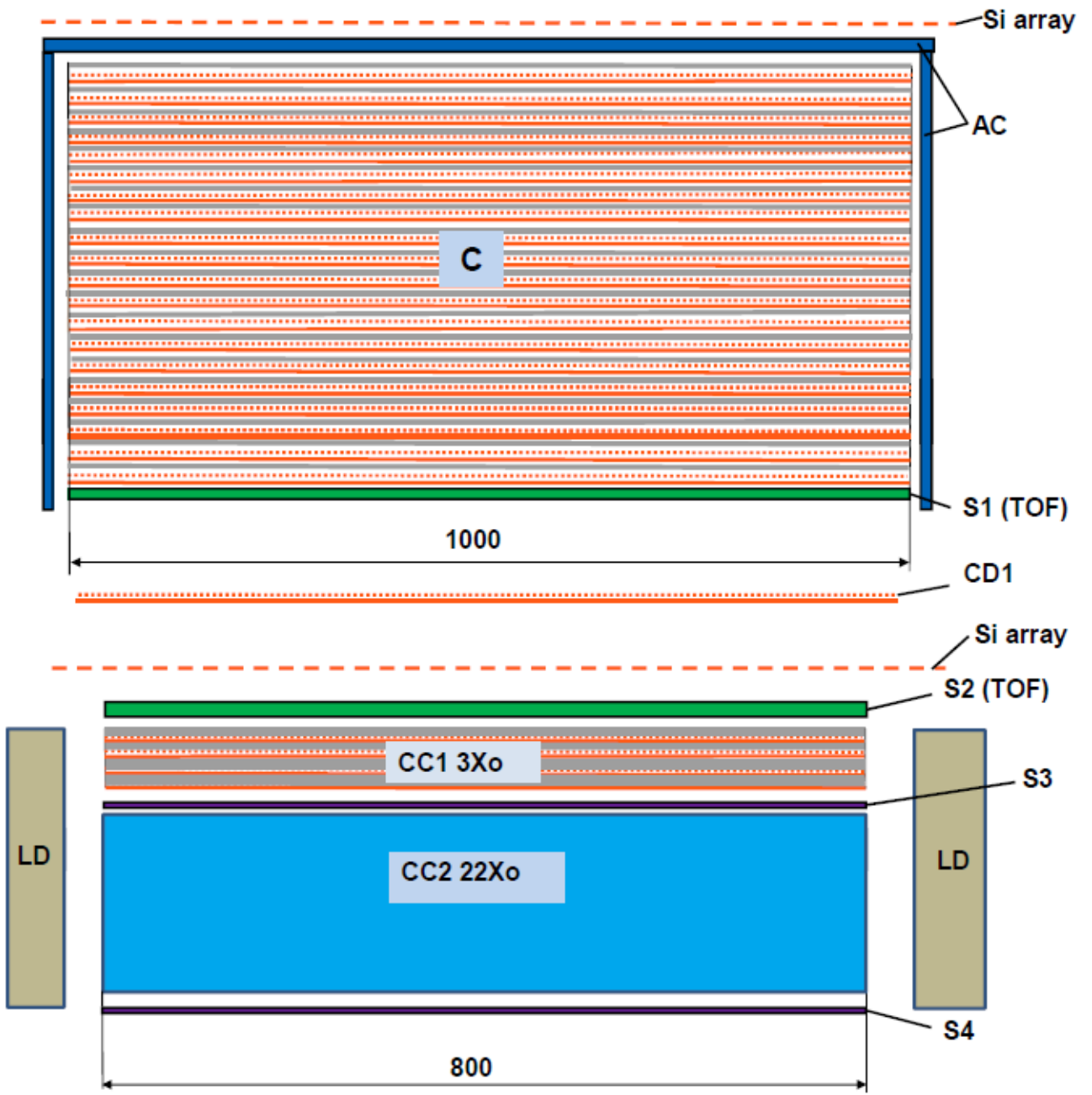

Fig. 1.

*Table 1. A comparison of basic parameters of space-based and ground-based instruments*

| | SPACED-BASED | | | | | GROUND-BASED | | | |
|---|---|---|---|---|---|---|---|---|---|
| | EGRET | AGILE | Fermi | CALET | **GAMMA-400** | H.E.S.S. | MAGIC | VERITAS | CTA |
| Energy range, GeV | 0.03-30 | 0.03-50 | 0.1-300 | 10-10000 | **0.1-3000** | >100 | >50 | >100 | >10 |
| Angular resolution, deg ($E_\gamma > 100$ GeV) | 0.2 $E_\gamma \sim 0.5$ GeV | 0.1 $E_\gamma \sim 1$ GeV | 0.1 | 0.1 | **~0.01** | 0.1 | 0.1 | 0.1 | 0.1 |
| Energy resolution, % ($E_\gamma > 100$ GeV) | 15 $E_\gamma \sim 0.5$ GeV | 50 $E_\gamma \sim 1$ GeV | 10 | 2 | **~1** | 15 | 20 | 15 | 15 |